\documentclass{amsart}

\usepackage{algorithm}
\usepackage{algpseudocode}
\usepackage{amsmath}
\usepackage{amssymb}
\usepackage{graphics}
\usepackage{epsfig}
\usepackage{graphicx}
\usepackage{multirow}
\usepackage{color}

\usepackage[caption=false,font=footnotesize]{subfig}

\newcommand{\AR}{\texttt{AR}}
\newcommand{\FR}{\texttt{FR}}
\newcommand{\CW}{\texttt{CW}}
\newcommand{\PW}{\texttt{PW}}

\newcommand{\ud}{\textup{d}}
\newcommand{\CLFNSA}{\texttt{CLF\_N\_OC}}
\newcommand{\CLFOC}{\texttt{CLF\_O\_C}}
\newcommand{\1}{\textsf{1}}
\newcommand{\2}{\textsf{2}}
\newcommand{\3}{\textsf{3}}
\newcommand{\NOR}{\texttt{N}}
\newcommand{\OSA}{\texttt{O}}
\newcommand{\CSA}{\texttt{C}}
\newcommand{\SA}{\texttt{OC}}
\newcommand{\sNOR}{\textsf{NOR}}
\newcommand{\sOSA}{\textsf{OSA}}
\newcommand{\sCSA}{\textsf{CSA}}
\newcommand{\ARtho}{\AR_{\texttt{tho}}}
\newcommand{\FRtho}{\FR_{\texttt{tho}}}
\newcommand{\ARabd}{\AR_{\texttt{abd}}}
\newcommand{\FRabd}{\FR_{\texttt{abd}}}
\newcommand{\Ytho}{Y_{\text{tho}}}
\newcommand{\Yabd}{Y_{\text{abd}}}
\newcommand{\Atho}{\widetilde{A}_{\text{tho}}}
\newcommand{\Aabd}{\widetilde{A}_{\text{abd}}}
\newcommand{\gtho}{g_{\text{tho}}}
\newcommand{\gabd}{g_{\text{abd}}}
\newcommand{\Cov}{\texttt{Cov}}

\newcommand{\RR}{\mathbb{R}}

\title[Sleep Apnea Detection via Piezo-Electric Bands]{Sleep Apnea Detection Based on Thoracic and Abdominal Movement Signals of Wearable Piezo-Electric Bands}

\author[Y.-Y. Lin]{Yin-Yan~Lin}
\address{Department of Electrical engineering, National Tsing-Hua University, Taiwan.} 
\email{sherry22110@gmail.com}

\author[H.-T. Wu]{Hau-Tieng~Wu}
\address{Department of Mathematics, University of Toronto}
\email{hauwu@math.toronto.edu}

\author[C.-A. Hsu]{Chi-An~Hsu}
\address{Department of Electrical engineering, National Tsing-Hua University, Taiwan.} 

\author[P.-C. Huang]{Po-Chiun~Huang}
\address{Department of Electrical engineering, National Tsing-Hua University, Taiwan.} 
\email{pchuang@ee.nthu.edu.tw}

\author[Y.-H. Huang]{Yuan-Hao~Huang}
\address{Department of Electrical engineering and Institute of Communications Engineering, National Tsing-Hua University, Taiwan}
\email{yhhuang@ee.nthu.edu.tw}

\author[Y.-L. Lo]{Yu-Lun~Lo}
\address{Department of Thoracic Medicine, Healthcare Center, Chang Gung Memorial Hospital, Chang Gung University, School of Medicine, Taipei, Taiwan}
\email{loyulun@hotmail.com}

\begin{document}

\begin{abstract}
Physiologically, the thoracic (THO) and abdominal (ABD) movement signals, captured using wearable piezo-electric bands, provide information about various types of apnea, including central sleep apnea (CSA) and obstructive sleep apnea (OSA). However, the use of piezo-electric wearables in detecting sleep apnea events has been seldom explored in the literature.
This study explored the possibility of identifying sleep apnea events, including OSA and CSA, by solely analyzing {one or both the THO and ABD signals. An adaptive non-harmonic model was introduced to model the THO and ABD signals, which allows us to design features for sleep apnea events. To confirm the suitability of the extracted features, a support vector machine was applied to classify three categories -- normal and hypopnea, OSA, and CSA. According to a database of} 34 subjects, the overall classification accuracies were on average $75.9\%\pm 11.7\%$ and $73.8\%\pm 4.4\%$, respectively, based on the cross validation. When the features determined from the THO and ABD signals were combined, the overall classification accuracy became $81.8\%\pm 9.4\%$. These features were applied for designing a state machine for online apnea event detection. Two event-by-event accuracy indices, S and I, were proposed for evaluating the performance {of the state machine. For the same database, the} S index was $84.01\%\pm 9.06\%$, and the I index was $77.21\%\pm 19.01\%$. The results indicate the considerable potential of applying the proposed algorithm to clinical examinations for both screening and homecare purposes.
\end{abstract}

\maketitle

\section{Introduction}
Sleep-disordered breathing (SDB) is a common disorder affecting approximately 14\% of adult men and 5\% of adult women \cite{Peppard_Young_Barnet_Palta_Hagen_Hla:2013}. Patients with SDB present frequent complete cessation of breathing and awakenings while sleeping at night, which leads to non-restorative sleep and hence excessive daytime sleepiness and fatigue. In addition, many evidences reveal that SDB is associated with several other diseases, such as hypertension \cite{Peppard_Young_Palta_Skatrud:2000}, heart disease \cite{Hamilton_Solin_Naughton:2004}, and stroke \cite{Bassetti_Aldrich:1999}. Moreover, SDB is responsible for several public disasters \cite{Colten_Altevogt:2006}. Although SDB has attracted a great amount of attention in the past decades, unfortunately, many patients with SDB are not appropriately diagnosed, \cite{Young_Evans_Finn_Palta:1997}. Therefore, screening subjects with SDB is a critical public health concern \cite{Phillipson:1993}.

The gold standard for SDB diagnosis is interpreting the multichannel signals recorded through polysomnography (PSG) \cite{Iber_Ancoli-Isreal_Chesson_Quan:2007}; However, this method has several limitations. {For example,} patients feel uncomfortable because numerous sensors might interfere with sleep; PSG measurement is highly labor-intensive and must be performed in a special environment, thus limiting its application to the whole population. {To resolve these limitations, in the past decades, numerous efforts have been made to identify a comfortable and easy-to-install wearable device for accurate automatic diagnosis. One frequently asked question is whether designing  a screening or monitoring system with few sensors, or even one sensor {(a level-IV monitoring system \cite{Ferber_Millman_Coppola:1994})}  is possible, without considering the electroencephalographic signal.
{For this purpose,} one specific challenge is evaluating the amount of clinical information that could be obtained from a single-channel sensor. 

The available information from some sensors {in PSG}, such as the nasal airflow signal, the electrocardiogram (ECG) signal, the oximeter signal, and  sound, has been well explored.
The nasal airflow signal is most directly related to the respiratory dynamics and has been widely investigated and applied to diagnose SDB \cite{Nakano_Tanigawa_Furukawa_Nishima:2007,Han_Shin_Jeong_Park:2008,Ragette_Wang_Weinreich_Teschler:2010,Rathnayake_Wood_Abeyratne_Hukin:2010,Koley_Dey:2013,Jiayi_Sanchez:2015}. 
The ECG signal that contains the perturbed physiological dynamics caused by apnea events, in particular, the heart rate variability, has also been extensively considered in the literature \cite{Tomas_Mietus_Peng_Gilmartin_Daly_Goldberger_Gottlieb:2007,Khandoker_Palaniswami_Karmakar:2009,Khandoker_Gubbi_Palaniswami:2009,Bsoul_Minn_Tamil:2011}.    }
The oximeter signal reflects the oxygen saturation and has been considered to identify sleep apnea events \cite{Burgos_Goni_Illarramendi_Bermudez:2010,Alvarez_Hornero_Marcos_Campo:2010}. 
Sound analysis provides another aspect of sleep apnea research. Specifically, whereas the aforementioned channels are associated with the physiological dynamics, snoring analysis provides more mechanical information about the upper airway structure \cite{Alshaer_Garcia_Radfar_Fernie_Bradley:2011,Lee_Yu_Lo_Chen_Wang_Cho_Ni_Chen_Fang_Huang_Li:2012}. 
{With the information available inside one single channel, integration of information from multiple channels has been beneficial for diagnosis} (e.g., oximeter combined with ECG \cite{Xie_Minn:2012} and oximeter combined with ECG and sound analysis \cite{Morillo_Ojeda_Foix_Jimenez:2010}). {We refer the reader to the review paper \cite{Alvarez-estevez2015} for a systematic summary.}

On the other hand, some commonly applied sensors in PSG have not been well explored, for example, the thoracic (THO) and abdominal (ABD) movement signals recorded by using the piezo sensors \cite{Nepal2002,Varady_Bongar_Benyo:2003,Al-Angari_Sahakian:2012,SLEEP_Lin2015}. Physiologically, these two channels contain information about sleep apnea and are used to classify various types of apnea, including central sleep apnea (CSA) and obstructive sleep apnea (OSA). However, these channels are not recommended as the first-line {sensors} {by American Academy of Sleep Medicine (AASM) \cite{Berry_Budhiraja_Gottlieb:2012}}, probably because of the instability of the piezo sensor {\cite{Folke2003}}. From the signal processing viewpoint, this type of signal is challenging because the instability nature is complicated by its non-stationarity nature, and suitable analysis tools are unavailable. This difficulty might explain its under-exploration.  
Based on the physiological understanding and practical clinical use of the THO and ABD signals, we hypothesized that in addition to providing apnea event information, {the essential information for classifying CSA and OSA events contained in the THO and ABD signals could be well extracted} when appropriate signal processing techniques are used. 
 
{To the best of our knowledge, few papers systematically explore the breathing information hidden in the THO and ABD signals. For example, the amplitude, energy and dominant frequency are considered in \cite{Nepal2002}, and an 100\% accuracy of classifying apnea and normal respiration is reported; the amplitude and phase are considered in \cite{Varady_Bongar_Benyo:2003}, and a 90.63\% averaged accuracy of classifying OSA, CSA, and normal respiration is reported; in \cite{Al-Angari_Sahakian:2012}, in addition to the amplitude and phase information, the phase relationship between ABD and THO is considered, and the best accuracy of classifying OSA and normal respiration is 69.9\% for the minute classification and 89\% for the subject classification. While the accuracies reported in \cite{Nepal2002} and \cite{Varady_Bongar_Benyo:2003} are high, only 266 and 189 episodes of respiratory signal were analyzed, respectively. In \cite{Al-Angari_Sahakian:2012}, only the OSA is considered in the study. Also, while the subject classification accuracy reported in \cite{Al-Angari_Sahakian:2012} is high, the AHI is used to classify the severity of subjects, and it is not clear if the predicted OSA events match the true OSA events.}

{In general, to properly explore the underlying information in the ABD and THO signals}, a model for the analysis is warranted. Although the respiratory signals, particularly the nasal airflow signal, have been extensively studied in the literature, no systematic model has been discussed. Based on the physiological understanding of the respiratory signal, a model could be designed to justify how and why an algorithm is applied to extract the features. Based on this model, the inevitable noise in the recorded signals can be more easily handled. 
Furthermore, to properly evaluate the performance of the algorithm, a proper accuracy metric is inevitable. The commonly applied apnea-hypopnea index (AHI) {and} apnea index (AI) 
{are not directly suitable for this purpose since only the event number is taken into account in the index. However, to evaluate the accuracy of the detected events by an algorithm, we need to perform} the {\it event-by-event} evaluation;
that is, a detected event should significantly overlap with a true event determined by the sleep expert.

{In this study, we applied a recently developed adaptive non-harmonic model and designed five dynamical features by considering the physiological facts {and properties of the piezo sensor}, to explore the underlying information hidden in the ABD or THO signal. 
{Specifically, we apply sychrosqueezing transform (SST) algorithm to reduce the influence of the inevitable noise in the recorded ABD and THO signals. Due to the instability of the piezo sensor, we consider the \textit{amplitude ratio}, instead of the amplitude, as a new feature; based on the nature of the piezo sensor, the \textit{frequency ratio} is proposed as another new feature; the \textit{covariance} of ABD and THO is considered as an auxiliary feature if both ABD and THO signals are used in the analysis.}
Furthermore, a support vector machine (SVM) \cite{Scholkopf_Smola:2002,Al-Angari_Sahakian:2012} was applied to classify the extracted features. Based on the classification results, we proposed an on-line algorithm for sleep apnea event detection {based on a newly designed finite state machine classifier}. The obtained result {of the finite state machine classifier} was justified by using two novel assessment indices on an event-by-event basis.  
{The} results support that with suitable signal processing, either ABD or THO has the potential to serve as a comfortable and easy-to-install SDB detection instrument. {A preliminary study of this work was reported in the conference \cite{SLEEP_Lin2015}.}} 

The remainder of this paper is organized as follows. Section \ref{sec:Mdl_and_Algo} introduces the physiological and medical background of SDB and the adaptive non-harmonic model to quantify the respiratory signals, including ABD and THO. 
Section \ref{sec:Matrl_and_Method} details the features designed for ABD and THO signals and the SVM classifier. 
Section \ref{Section:Result:SVM} reports the study material and SVM classification results.
Section \ref{subsection:StateMachine} applies the designed features and established SVM classifiers to design a state machine for potential online prediction purpose. In addition, two accuracy assessment indices for evaluating the event-by-event detection are introduced. 
Section \ref{sec:Disc} discusses our findings and concludes this paper.

\section{Physiological background and model}\label{sec:Mdl_and_Algo}

This section first summarizes the essential physiological background of sleep apnea and its phenomenological behavior observed from the ABD and THO signals. Based on these physiological facts, we proposed the {\it adaptive non-harmonic model} to quantify these signals, which serves as a basis for designing our features.

\subsection{Physiological background}
{Generally, SDB comprises the following five types: sleep apnea, sleep hypopnea (HYP), respiratory effort-related arousal (RERA), sleep hypoventilation, and Cheyne-Stokes breathing. These events are distinguished by the cause of the shallowing or cessation of breathing. OSA, obstructive sleep HYP, and RERA are caused by a completely or partially blocked upper airway caused by decreasing muscle tone, in particular the muscles surrounding the upper airway. During these obstructive events, the subject attempts to breathe; therefore, unusual movements are observed in the THO and ABD signals. A significant pattern of obstructive event is a paradoxical movement, during which the direction of the {thoracic} movement is opposite to that of the {abdominal} movement. However, the paradoxical movement does not occur in all obstructive events. 

CSA, central sleep HYP, and Cheyne-Stoke breathing are completely distinct from the obstructive events in terms of both the physiological mechanism and breathing pattern. In these types, the respiratory control center in the brain is imbalanced in patients with central sleep breathing disorder -- the carbon dioxide level in the blood is imbalanced and the neurological feedback mechanism that monitors the carbon dioxide level does not  function properly, causing difficulty of maintaining the respiratory rate. By contrast to the obstructive events, the subject with central sleep breathing disorder completely or partially stops breathing and makes little or no effort to breathe during those central events. Thus, the {thoracic} and {abdominal} movements decrease or disappear during those central events. 

Mixed sleep apnea (MSA) is an event characterized by complete cessation of respiration without {thoracic} and {abdominal} movements initially. The {thoracic} and {abdominal} movements gradually appear before the end of the event. Because the breathing pattern exhibits the characteristics of both OSA and CSA, the event is termed as MSA. Clinically, MSA has been considered as a variant of OSA, but not CSA \cite{AASMtask:1999}; therefore, we considered MSA events as OSA events in our study.

By contrast to obstructive, central or mixed sleep disordered events, sleep hypoventilation is characterized by abnormal ventilation, but not apnea or HYP, and gas exchange that considerable aggravates or may only occur during sleep. These abnormalities result in hypercapnea and are sometimes associated with hypoxemia. Sleep hypoventilation may be primarily related to blunted chemo-responsiveness or be comorbid with particular medical conditions that cause the impairment of gas exchange. We exclude patients with sleep hypoventilation in this study.

Quantitatively, an apnea event (OSA or CSA) is identified when the airflow breathing amplitude decreases more than 90\% for a duration ranging from 10 to 120 seconds, whereas an HYP event is identified when either of the following two conditions holds: (1) the airflow breathing amplitude decreases more than 30\% of the pre-event baseline with $\geq 4$\% oxygen desaturation; (2) the airflow breathing amplitude decrease more than 50\% of the pre-event baseline with $\geq 3$\% oxygen desaturation or with an arousal for a duration ranging from 10 to 120 seconds, but does not fulfill the criteria for apnea. In this study, we followed the {AASM} 2007 \cite{Iber_Ancoli-Isreal_Chesson_Quan:2007} and did not classify an HYP event as a central or obstructive in nature, although the {AASM} updated the scoring criteria for sleep disordered breathing events in 2012 \cite{Berry_Budhiraja_Gottlieb:2012}. The 2012 scoring rule \cite{Berry_Budhiraja_Gottlieb:2012} for sleep apnea is identical to that in 2007, whereas the HYP rule is modified as ``a HYP event is identified when the airflow breathing amplitude decreases over 30\% of the pre-event baseline with $\geq3$\% oxygen desaturation or with an arousal''. This modification, however, {does not influence} the data analysis in this study, {as we focus on the apnea events classification}.

To quantify the severity of apnea in clinics, the AHI is applied and is defined by dividing the number of apnea and hypopnea events by the number of sleep hours. The AI and hypopnea index (HI) are defined in a similar manner.} Based on the AHI index, the severity of apnea in patients is classified as normal (AHI$\leq$ 5), mild (5 $<$AHI$\leq$ 15), moderate (15 $<$AHI$\leq$ 30) and severe (AHI$>$ 30).

\subsection{Adaptive non-Harmonic Model}

Driven by the physiological facts discussed in the previous section, we proposed an adaptive non-harmonic model to quantify the THO and ABD movement signals. Although the respiratory activity is oscillatory, it is ``irregular''. First, sleep is a global and systematic behavior that involves all body parts. In particular, even during normal sleep, the muscular atonia and low amplitude electromyography (EMG) are intimately related to the sleep cycle \cite{Lee-Chiong:2008}, which leads to significant changes in the breathing rate and pattern \cite{Wu_Talmon_Lo:2015}. In subjects with sleep apnea, the condition is becoming more complicated. Moreover, in addition to the time-varying frequency caused by the sleep cycle, the amplitude of the ABD and THO signals might vary and even become zero during a central apnea event. In addition, although not significant, the heart beats contribute to the ABD and THO movement, and this movement becomes dominant during a CSA event. 
To quantify the ABD and THO movement signals, we considered the following model.

We first introduce the intrinsic mode type (IMT) function.
Fix $0\leq \epsilon\ll 1$. Consider the set $\mathcal{C}_{\epsilon}$ that consists of differentiable and bounded functions $g(t) = A (t)s(\phi(t))$, where $A$, $s$, and $\phi$ satisfy the following conditions: (1) $A$ is positive, continuous, and bounded and its first-order derivative is continuous; (2) $\phi$ increases monotonically and its first two derivatives are continuous; (3) the absolute values of the first-order derivative of $A$ and second-order derivative of $\phi$ at time $t$ are bounded by $\epsilon\phi'(t)$, for all $t\in\RR$; and (4) $s:[0,1]\to \RR$ is a continuous $1$-periodic function with unit $L^2$ norm such that $|\widehat{s}_\ell(k)|\leq \delta |\widehat{s}_\ell(1)|$ for all $k\neq 1$, where $\delta\geq 0$ is a small parameter. 
The $1$-periodic function $s(\cdot)$ is called the {\it wave shape function}, which describes the mechanism of signal oscillation over one oscillation. The theoretical details of the wave shape function are described in \cite{Wu:2013,lin2016waveshape}. 
The positive function $A(t)$ describes the amplitude of the oscillation at time $t$, and the positive function $\phi'(t)$ describes the speed of the oscillation at time $t$. We consider $A(t)>0$ to be the {\it amplitude modulation} (AM) and $\phi'(t)>0$ to be the {\it instantaneous frequency} (IF) of an oscillatory function $g(t)$. Note that the IF and AM are always positive, but usually not constant. The conditions $|A'(t)|\leq \epsilon \phi'(t)$ and $|\phi''(t)|\leq \epsilon\phi'(t)$ force the signal to locally behave like a harmonic function. We consider $g(t)$ that satisfies the aforementioned conditions as an IMT function.  The theoretical details of an IMT function are described in \cite{Daubechies_Lu_Wu:2011} .

We introduced AM and IF  to quantify the possible apnea events as well as the time-varying breathing rate. The wave shape function was introduced to capture the non-harmonic nature of a breathing cycle. For example, in the general respiratory activity, the inspiratory period is shorter than the expiratory period, which can not be captured by a cosine function. 

In most real data, an oscillatory signal might be composed of more than one IMTs. For example, in the THO and ABD movement signals, in addition to the respiratory movement, the oscillatory movement induced by the heart beats is also recorded. Thus, we proposed to model the THO and ABD movement signals by using the following {\it adaptive non-harmonic model}. Fix constants $0\leq \epsilon\ll 1$ and $d>0$. Consider the set $\mathcal{C}_{\epsilon,d}$ that consists of differentiable and bounded functions such that $G(t) = \sum_{\ell=1}^K g_\ell(t)$, where $K$ is finite and $g_k(t)=A_\ell(t)s_\ell(2\pi\phi_\ell(t))\in \mathcal{C}_{\epsilon}$; when $K>1$, $\phi_{\ell+1}'(t)-\phi'_\ell(t)>d$ for all $\ell=1,\ldots,K-1$ is satisfied. We define a function $G$ satisfying the aforementioned conditions as an adaptive non-harmonic model.

Therefore, the ABD and THO signals are modeled by using the adaptive non-harmonic model with $K=2$. In this case, $g_1$ is associated with the respiratory movement with a lower IF of approximately $0.4$ Hz and higher amplitude, whereas $g_2$ is associated with the movement induced by the heart beats with a higher IF of approximately $1.2$ Hz and lower amplitude. 

However, because the recorded signal is contaminated by noise in practice, we consider the final phenomenological model to describe the recorded ABD and THO signals.
\begin{equation}\label{model:adaptiveNonHarmonic}
\left\{
\begin{array}{l}
\Ytho(t) = \gtho(t)+\sigma_{\text{tho}}(t)\Phi_{\text{tho}}(t)\,;\\
\Yabd(t) = \gabd(t)+\sigma_{\text{abd}}(t)\Phi_{\text{abd}}(t),
\end{array}
\right.
\end{equation}
where { $\Ytho$ is the recorded THO signal, $g_{\text{tho}}(t)=A_{\text{tho}}(t)s_{\text{tho}}(2\pi\phi_{\text{tho}}(t))+A_{\text{hb}}(t)s_{\text{hb}}(2\pi\phi_{\text{hb}}(t))$ is the clean signal in $\mathcal{C}_{\epsilon,d}$ containing the THO movement $A_{\text{tho}}(t)s_{\text{tho}}(2\pi\phi_{\text{tho}}(t))\in \mathcal{C}_{\epsilon}$ and the movement induced by the heart beats $A_{\text{hb}}(t)s_{\text{hb}}(2\pi\phi_{\text{hb}}(t))\in \mathcal{C}_{\epsilon}$, $\Phi_{\text{tho}}$ is the stationary stochastic random process, and $\sigma_{\text{tho}}(t)$ is a smooth and slowly varying function. Here $\sigma_{\text{tho}}(t)\Phi_{\text{tho}}(t)$} models the possible non-stationarity in the measurement noise. A similar interpretation holds for $\Yabd$, $\gabd$, $\sigma_{\text{abd}}$ and $\Phi_{\text{abd}}$.   The adaptive non-harmonic model has been further discussed in \cite{Daubechies_Lu_Wu:2011,Chen_Cheng_Wu:2014,Wu:2013}.

\section{Feature Design and Extraction and Classification}\label{sec:Matrl_and_Method}

In this section, we propose features for the ABD and THO signals, and detail the algorithms to extract them. 
On the basis of the physiological knowledge, we proposed two features, $\ARtho$ and  $\FRtho$, for the THO signal, and two features, $\ARabd$ and $\FRabd$, for the ABD signal. Before extracting the features from each window, the online sychrosqueezing transform (SST) algorithm was applied to reduce the influence of the inevitable noise in the recorded ABD and THO signals.
The proposed features were then fed into the SVM to obtain a two-layer binary SVM classifier. The first layer classifier, $\CLFNSA$ ($\texttt{CLF}$ denotes a classifier), classifies non-apnea event (denoted as NOR, i.e., OSA or CSA does not occur) and apnea events (OSA or CSA occurs); the second layer classifier, $\CLFOC$, classifies apnea events to OSA and CSA. To avoid the over-fitting problem, the cross validation was applied to evaluate the proposed features and the classification accuracy of the SVM model. We did not attempt to classify NOR into hypopnea and non-hypopnea because of the limitation of the proposed information available in the ABD and THO signals.

\subsection{Online adaptive denoise}
Here we summarize an online adaptive denoise algorithm to stabilize the possible noise in the ABD and THO signals. Consider the recorded ABD signal $\Yabd(t)=g_{\text{abd}}(t)+\sigma_{\text{abd}}(t) \Phi_{\text{abd}}$. The same algorithm can be applied to the THO signal. Let $g_{\text{abd}}(t)=A_{1}(t)s_{1}(\phi_{1}(t))+A_{2}(t)s_{2}(\phi_{2}(t))$, where $A_{1}(t)s_{1}(\phi_{1}(t))$ represents the abdominal movement induced by respiration and $A_{2}(t)s_{2}(\phi_{2}(t))$ represents the abdominal movement induced by the heart beats. Here, $A_{2}(t)$ is much smaller than $A_{1}(t)$. During an apnea event, $A_{1}(t)$ diminishes or becomes zero whereas $A_{2}(t)$ remains unchanged. Although these facts allow us to distinguish various types of apnea events, obtaining $A_{1}(t)$ from $g_{\text{abd}}(t)$ directly when noise exists is not an easy task, particularly when the noise is non-stationary; for example, when the subject moves. To reliably obtain $A_1$ as the feature for apnea at each time point, we applied the online SST. 

Consider $h(x)=e^{-x^2/2\sigma^2}$, where $\sigma>0$, as a window. In this study, we fixed $\sigma=2$. At each time point $t_0$, we estimated $A_1(t)$ by using the following three steps. First, for $t\in [t_0-\sigma,t_0+\sigma]$, we evaluated the Fourier transform of the signal ${\Yabd}(x)h(x-t)$, which is denoted as $V_{\Yabd}(t,\xi)$. Clearly, $V_{\Yabd}(t,\xi)$ is a well-known short time Fourier transform at time $t$. Second, we calculated the reassignment rule by $\omega_{\Yabd}(t,\xi)=\frac{-i\partial_tV_{\Yabd}(t_0,\xi)}{2\pi V_{\Yabd}(t,\xi)}$ when $|V_{\Yabd}(t,\xi)|\neq 0$, where $i$ is the imaginary unit and $\partial_t$ is the partial derivative with related to $t$, and $\omega_{\Yabd}(t,\xi)=-\infty$ otherwise. 
Third, the AM of the ABD signal at time $t_0$ was evaluated by
\begin{align}
\Aabd(t_0):=\left|\int_{\mathfrak{W}_{t_0}}S_{\Yabd}(t_0,\xi)\ud \xi \right|,\label{algorithm:sst:amplitude}
\end{align}
where $\mathfrak{W}_{t_0}:=\{\xi:~|\phi'(t_0)-\xi|\leq \epsilon\}$,
\begin{equation}\label{alogithm:sst:formula}
S_{\Yabd}(t_0,\xi):=\int_{\mathfrak{Q}_{t_0}}\delta\left(|\omega_{\Yabd}(t_0,\eta)-\xi|\right)V_{\Yabd}(t_0,\eta)\ud \eta,
\end{equation}
$\mathfrak{Q}_{t_0}:=\{\eta:~|V_{\Yabd}(t_0,\eta)|\geq 10^{-8}\}$, and $\phi_1'(t_0)$ is the estimated instantaneous frequency at $t_0$. Similarly, we could estimate the AM of the thoracic movement, denoted by $\Atho(t)$. Details of the algorithm and theory beyond the algorithm, in particular, the online adaptive denoise, are presented in \cite{Daubechies_Lu_Wu:2011,Chen_Cheng_Wu:2014,Chui_Lin_Wu:2014}.

\subsection{Feature Design}
The physiological background of sleep apnea provided abundant information to guide us in selecting features for the analysis. In addition to the physiological background, we considered the piezo-electric sensor feature to design a good feature.

\begin{figure}[t]
\centering
\includegraphics[width=.85\textwidth]{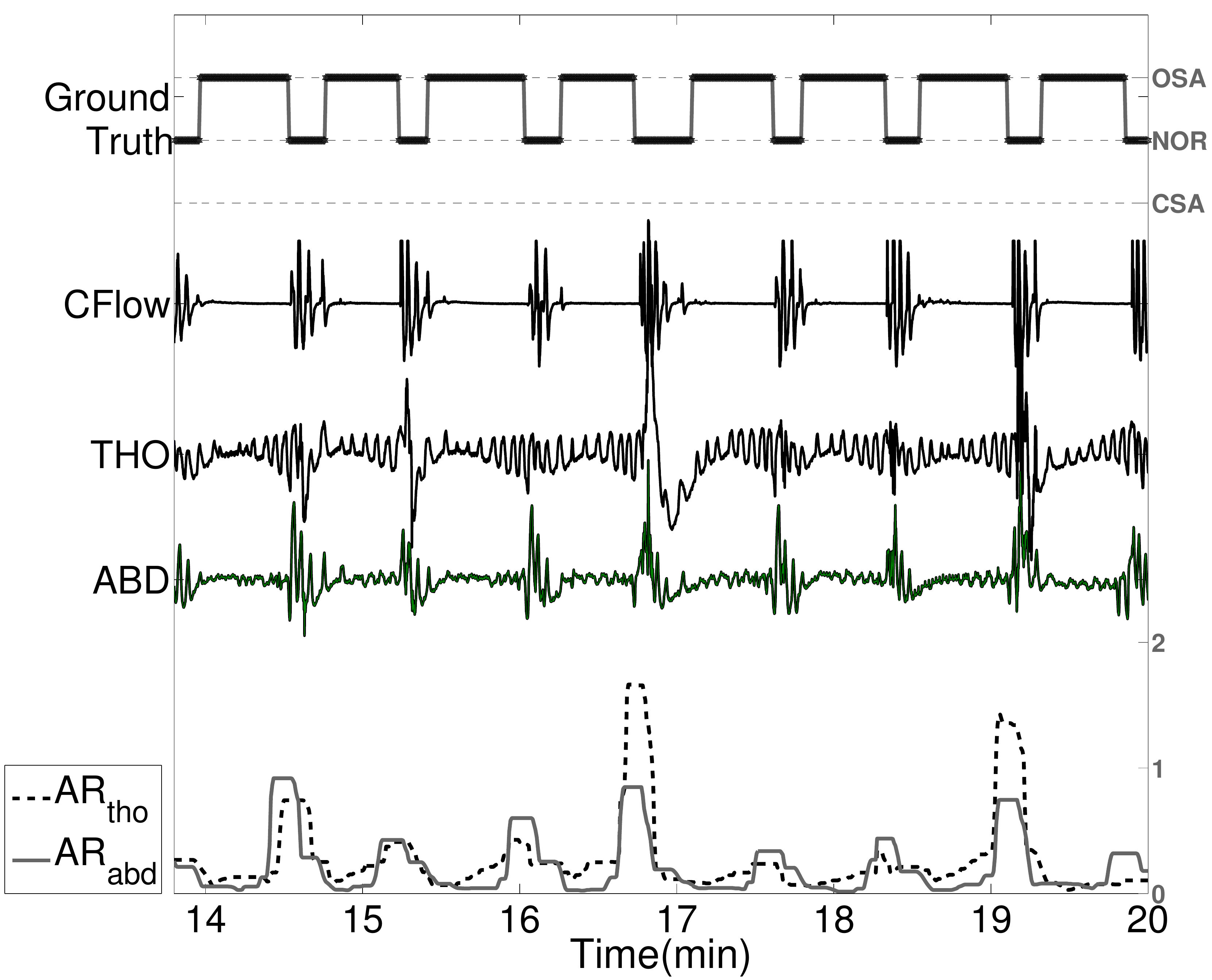} 
\caption{An illustration of amplitude ratio (AR) feature from a subject with obstructive sleep apneas (OSA). The signal duration is from the $14$-th min to the $20$-th min after the subject falls asleep. The top panel shows the sleep apnea state, including OSA, central sleep apnea (CSA), and non-apnea (NOR) status, evaluated by the sleep expert. The second panel shows the nasal flow (CFlow) signal. Clearly, intermittent apnea events occur. The third and fourth panels display the thoracic (THO) and abdominal (ABD) movement signals. When the CFlow signal was flat, we could observe oscillations in ABD and THO, reflecting the efforts of the subject to breathe. The bottom panel shows the AR features extracted from the ABD and THO signals. Clearly, when apnea events occur, the AR features significantly decrease.} 
\label{fig:AR_wf} 
\end{figure}

First, when the amplitudes of $\gtho(t)$ and $\gabd(t)$ decrease during a hypopnea or an apnea event, {the apnea or hypopnea is not the only resource for amplitude variation.} The body movement during sleep also contributes to the amplitude variation because of the time-varying contact between the body and piezo-electric bands. Thus, the \textit{absolute} amplitude is not a suitable feature for apnea event detection. Alternatively, because body movement is not frequent, we considered the ratio of amplitudes during consecutive time points {to suppress the influence caused by changes in posture}. This ratio is small during an apnea event. See Figure \ref{fig:AR_wf} for an example. 

\begin{figure}[t]
\centering
\includegraphics[width=.85\textwidth]{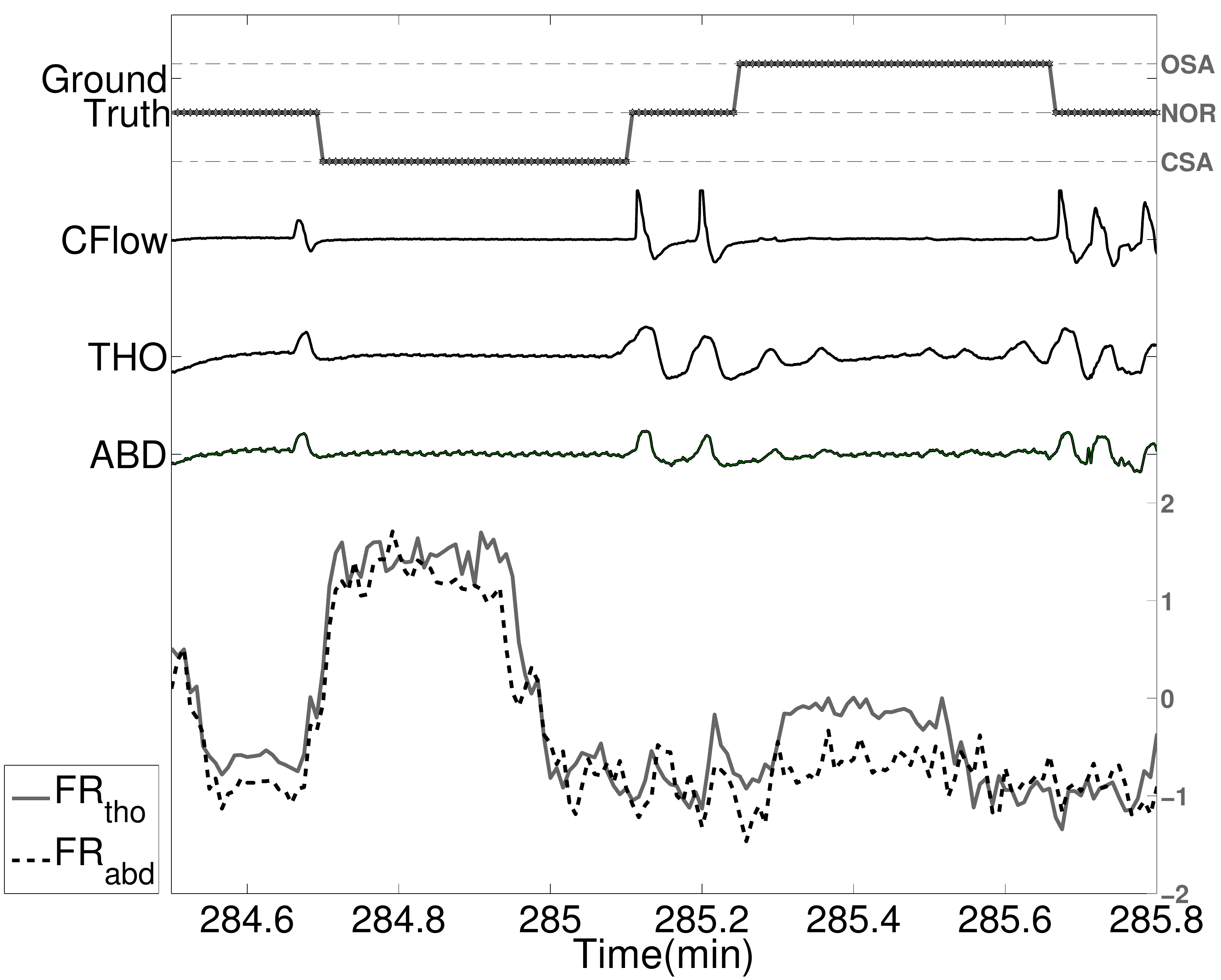} 
\caption{An illustration of frequency ratio (FR) feature from a subject with obstructive sleep apneas (OSA) and central sleep apnea (CSA). The signal duration is from the $284.4$-th min to the $285.8$-th min after the subject falls asleep. The top panel shows the sleep apnea state, including OSA, CSA, and non-apnea (NOR) status, evaluated by the sleep expert. Second panel shows the nasal flow (Cflow) signal. Two clear apnea events can be observed around the $284.8$-th min and the $285.5$-th min. The third and fourth panels display the thoracic (THO) and abdominal (ABD) movement signals. We could see a complete cessation of respiratory movement around the $284.8$-th min, which indicates the event of CSA. Note that during this period the heart beats could be detected and is represented as a small and regular oscillation. The bottom panel shows the FR features. Note that there is an obvious surge of the FR features during the CSA event.} 
\label{fig:FR_wf} 
\end{figure}

Second, during the CSA events, when the respiratory activity stops completely, we could observe the high frequency oscillation induced by the heart beats in $\gtho(t)$ and $\gabd(t)$. Thus, during the CSA events, the energy on the higher frequency band is dominant. See Figure \ref{fig:FR_wf} for an example. Based on these physiological facts, we considered two sets of features: {\it amplitude ratio} (AR) and {\it frequency ratio} (FR). Third, {the paradoxical movement is a crucial feature for OSA, although it does not always occur during the OSA event. To quantify the paradoxical movement, we could consider the correlation between the ABD and THO movements. The correlation information between two channels has been successfully used in other sleep apnea detection algorithms, for example, \cite{kumar2005analysis,Coleman_Roffwarg_Kennedy:1982,Simmons01081984}.}

\subsection{Feature Extraction}

To extract the features, we segmented the signals into overlapping windows of $10$-s duration with $9.5$-s overlap. These windows are called the {\it current window}s (CW), which provide potential information about the apnea event. The $n$-th  CW is denoted as $\CW(n)\subset \RR$. Note that $\CW(n)\neq\CW(m)$ when $n\neq m$. For AR features, we considered the closest window of $60$-s duration in which no apnea was reported by the sleep expert. We termed this window as the {\it pre-window} (PW), which contains the baseline information for the definition of AR. The $n$-th PW associated with the $n$-th CW is denoted as $\PW(n)\subset \RR$. Note that $\PW(n)$ might be the same as $\PW(m)$ when $n\neq m$, particularly during the apnea event. To be more precise, the PW is fixed when the CW was moving forward if the CW was annotated as a sleep apnea event by the sleep expert. 
This result is illustrated in Figure \ref{fig:algoflow} (top panel). 

{The selection of CW and PW durations was based} on the criteria by which sleep experts annotated the sleep apnea; in practice, they compare the amplitudes of the oral-nasal flow and THO and ABD movement signals in the current $10$ s with the average of those in the previous $120$ s to determine an event. In our application, however, we did not use the oral-nasal flow for comparison. Moreover, the average information coming solely from the previous $120$-s ABD and THO movements might comprise other events or abnormal signals and hence might downgrade the information in the extracted features. Therefore, we selected $60$-s duration as our PW.

\subsubsection{Respiratory amplitude ratio}
Due to the noise, to extract AR features in a robust way, we considered $\Atho(t)$ and $\Aabd(t)$ estimated by the online SST (\ref{algorithm:sst:amplitude}) instead of the original signal.
The AR of the $n$-th CW is derived by
\begin{equation}\label{eq_AR}
\begin{split}
&\ARtho(n)=\frac{Q_{95}(\Atho(t)\chi_{\CW(n)})}{Q_{95}(\Atho(t)\chi_{\PW(n)})}\\
&\ARabd(n)=\frac{Q_{95}(\Aabd(t)\chi_{\CW(n)})}{Q_{95}(\Aabd(t)\chi_{\PW(n)})},
\end{split}
\end{equation} 
where $\ARtho(n)$ and $\ARabd(n)$ represent the ARs of the THO and ABD signals, respectively, over the $n$-th window, $\chi$ is the indicator function, and $Q_{95}$ represents $95\%$ quantile of the given function. Here we selected the $95\%$ quantile instead of the maximum to avoid outliers caused by the noise in the signals. One immediate benefit of using the ratio, rather than the absolute value, {is the alleviation of common drawbacks of the piezo sensor like the ``trapping artifact'' -- the elastic belt tension might be distorted by the movement \cite[p 662]{Butkov_Lee-Chiong:2007}.} In addition, it helps automatic removal of the inter-individual discrepancy and time-varying amplitude induced by other physiological facts. Certainly, although various subjects have varying physiological profiles such as tidal volume and breathing rate under various sleep stages during the apnea event, the AR is still small.

\subsubsection{Respiratory Frequency Ratio}

\begin{figure}[t]
\centering
\includegraphics[width=.85\textwidth]{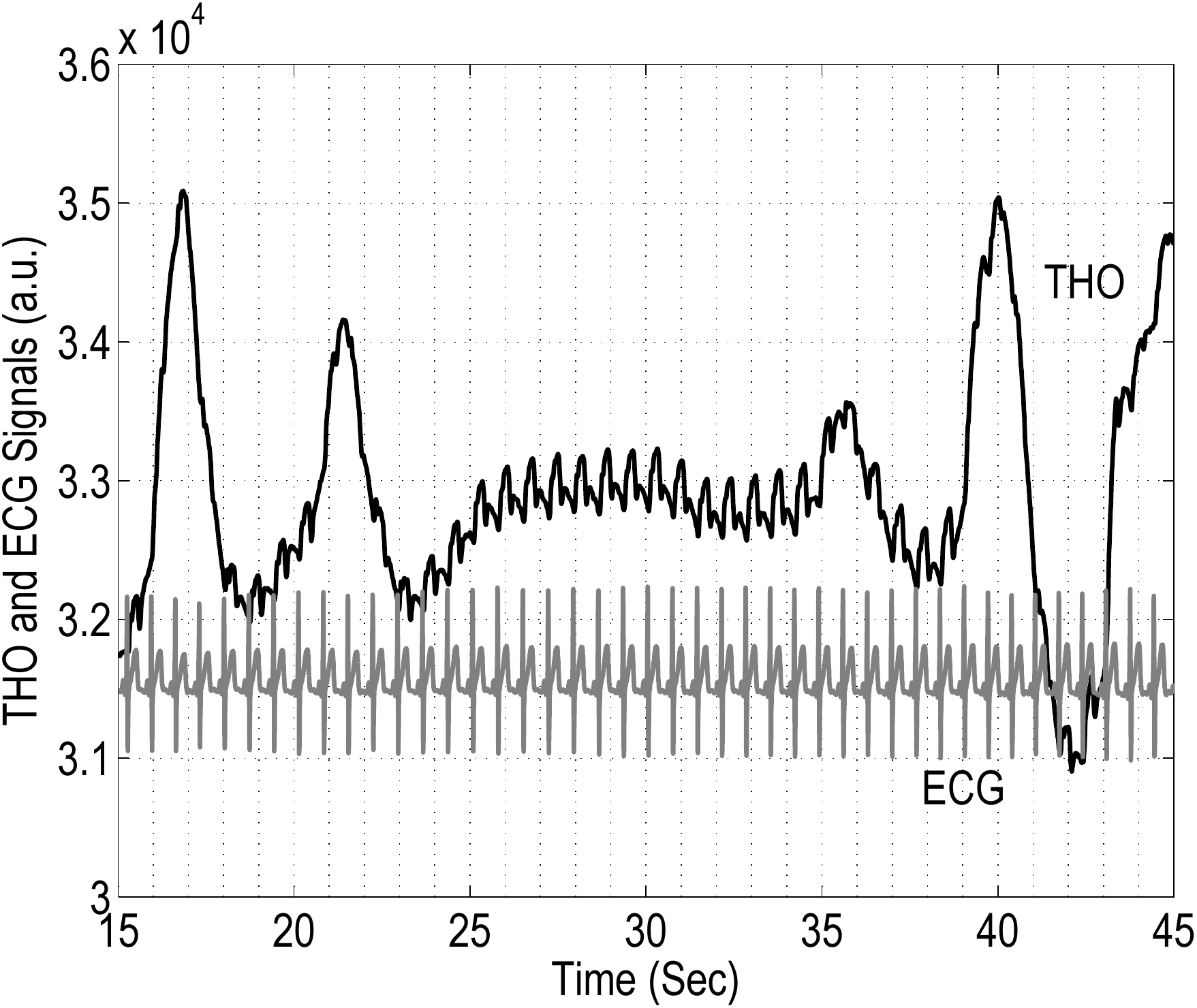} 
\caption{The THO signal versus ECG signal during a CSA event. The oscillation in the THO signal is mainly dominated by the {cardiac activity} when a CSA event occurs. {This is commonly known as the cardiogenic artifact.}} 
\label{fig:ECGvsTHO} 
\end{figure}

The FR of the $n$-th CW is derived by
\begin{equation}\label{eq_FR}
\begin{split}
&\FRtho\text{(n)}=\log_{10}\left(\frac{\int_{0.8}^{1.5} |\mathcal{F}(\Ytho(t)\chi_{\CW(n)})(\xi)|^2\ud \xi}{\int_{0.1}^{0.8} |\mathcal{F}(\Ytho(t)\chi_{\CW(n)})(\xi)|^2\ud \xi} \right) \\
&\FRabd\text{(n)}=\log_{10}\left(\frac{\int_{0.8}^{1.5} |\mathcal{F}(\Yabd(t)\chi_{\CW(n)})(\xi)|^2\ud \xi}{\int_{0.1}^{0.8} |\mathcal{F}(\Yabd(t)\chi_{\CW(n)})(\xi)|^2\ud \xi}\right) ,
\end{split}
\end{equation} 
where $\FRtho(n)$ and $\FRabd(n)$ denote the $n$-th FRs of THO and ABD movement signals, respectively, and $\mathcal{F}$ represents the Fourier transform. Note that the main purpose of FR is to capture the possible CSA events. Specifically, when the respiratory activity ceases completely during a CSA event, $\Ytho(t)$ and $\Yabd(t)$ are composed mainly of the movement caused by the heart beats{, which is commonly considered as the cardiogenic artifact in the peizo sensor \cite[p 53]{Randerath_Sanner_Somers:2006}}.
Therefore, the integration in the dividend of (\ref{eq_FR}) ranges from 0.8 to 1.5 Hz, which is normally the range of the heart rate. { Figure~\ref{fig:ECGvsTHO} shows an illustration of the THO movement signal caused by the heart beats during a CSA event.}

Consequently, for the $n$-th CW, we obtained a vector $v(n)\in\RR^4$ composed of four features, $\ARtho(n)$, $\ARabd(n)$, $\FRtho(n)$ and $\FRabd(n)$. We termed $v(n)$ as the {\it respiratory activity index} of the $n$-th CW.

\subsubsection{Covariance between thoracic and abdominal movements as a feature}

In addition to the respiratory activity index, we quantified the paradoxical movement by using the covariance between the THO and ABD signals over $\CW$. The feature is called $\Cov$. Note that although the paradoxical movement is a crucial characteristic to define the OSA event, it is not considered as the main feature used to distinguish apnea events. Certainly, it does not occur in all OSA events because of the asynchronous phase caused by body movement. However, $\Cov$ could be used as an auxiliary feature to further confirm whether a given apnea event is obstructive.

\subsection{Ground truth by sleep experts}

We considered the respiratory activity scored by sleep experts according the AASM 2007 guideline \cite{Iber_Ancoli-Isreal_Chesson_Quan:2007} as the ground truth. We referred to the expert's score as the ``PSG state.''  
The PSG states are evaluated every $0.5$ s. Therefore, for a subject with a sleep record lasting for $N$ s, we obtained a time series, denoted as $s_{\text{PSG}}$, of length $2N-139$ with the range $\{\sNOR,\sOSA,\sCSA\}$, where 139 comes from the tail waveform of the PW and CW durations.

The respiratory activity from the $\ell$-th subject was classified into four groups, $\NOR$, $\OSA$, $\CSA$, and $\texttt{X}$, according to following rules:
\begin{enumerate}
\item If $s_{\text{PSG}}=\sNOR$ over $\CW(n)$, $v(n)$ is in the $\NOR$ group.
\item If $s_{\text{PSG}}=\sOSA$ over $\CW(n)$, $v(n)$ is in the $\OSA$ group.
\item If $s_{\text{PSG}}=\sCSA$ over $\CW(n)$, $v(n)$ is in the $\CSA$ group.
\item If $s_{\text{PSG}}$ contains more than one state, $v(n)$ is in the $\texttt{X}$ group.
\end{enumerate}
The $\texttt{X}$ group was considered as unknown and was excluded from the training process, whereas all $v(n)$ in $\texttt{X}$ are included in the testing process. 

\begin{figure}[t]
\centering
\includegraphics[width=.85\textwidth]{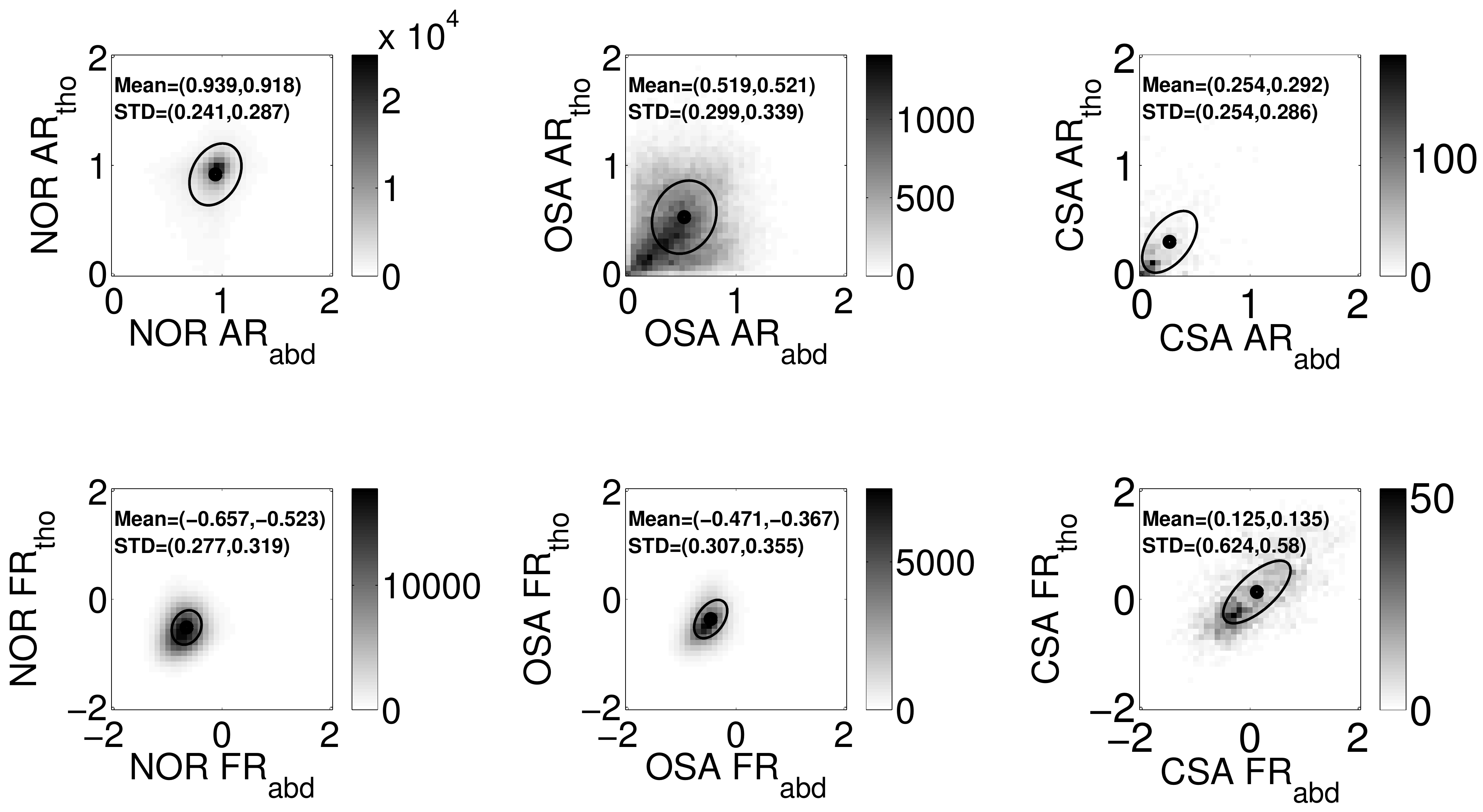} 
\caption{Feature distribution of amplitude ratio (AR) and frequency ratio (FR). The left, middle, and right columns display the distributions of AR features and FR features of the non-apnea ($\NOR$), obstructive sleep apnea ($\OSA$), and central sleep apnea ($\CSA$) groups, respectively, from all $34$ subjects as two dimensional histograms. 
The means of the AR and FR features are superimposed on the histogram as solid black dots, and the ellipsoids associated with the two principal directions of the covariance are also superimposed to enhance the visualization of the diversity the features. Note that in the $\CSA$ group, the mean of AR features slightly deviates from the visual center. This is caused by the dissemination of FR features. The mean and standard deviation (STD) of each feature are also shown on the plots.} 
\label{fig:ARFR_distr} 
\end{figure}

The distributions of $\ARtho$ versus $\ARabd$ and $\FRtho$ versus $\FRabd$ from the $\NOR$, $\OSA$, and $\CSA$ groups from the $\ell$-th subject are displayed in the top and bottom panels, respectively, in Figure \ref{fig:ARFR_distr} to evaluate the suitability of these features. The difference in the distributions of the respiratory activity indices from the $\NOR$, $\OSA$ and $\CSA$ groups can be visually observed. Note that $\FRtho$ and $\FRabd$ from the $\CSA$ group are distributed on the higher value region compared with those from the $\NOR$ and $\OSA$ groups. This phenomenon reflects the fact that during the CSA event, the movement induced by the heart beats is dominant. 

\subsection{Support Vector Machine as Classifier}\label{sec:SVM_training}

SVM has been widely applied in the sleep study literature \cite{Khandoker_Palaniswami_Karmakar:2009,Al-Angari_Sahakian:2012,Wu_Talmon_Lo:2015}. In a nutshell, a binary SVM classifier determines a hyperplane in the space separating the data set into two disjoint subsets, such that each subset lies in a different side of the hyperplane. According to the reproducing kernel Hilbert space theory, SVM is generalized to the {\it kernel SVM}, which facilitates classification with a nonlinear relationship. Technical details are further discussed in \cite{Scholkopf_Smola:2002}. For identifying the (possible) nonlinear relationship between various sleep apnea events, in this study, we applied the kernel SVM based on the standard radial based function as the kernel function. We applied two binary SVM classifiers, that is, the one-versus-all classification scheme \cite{Rifkin_Klautau:2004}, to achieve multiple group classification. Despite its simplicity, this scheme is highly effective. 
To prevent over fitting and validate the classification result, we applied the cross validation method. We ran the repeated random sub-sampling validation 25 times and reported the average; that is, we randomly separated the data into the training dataset and testing dataset -- the training dataset comprises randomly selected 80\% of the subjects and the remainder serves as the testing dataset. The trained classifier based on the training dataset was applied to predict the respiratory activity of the testing dataset. The classification accuracy in all subjects are reported as mean $\pm$ standard deviation unless otherwise specified.
 
For each subject, we trained two binary SVM classifiers from a given dataset based on the THO signal or THO and ABD signals. The first classifier distinguishes $\NOR$ from $\SA$ based on the respiratory activity indices, where $\SA:=\OSA\cup\CSA$. This classifier is denoted as $\CLFNSA$. The second classifier distinguishes $\CSA$ from $\OSA$ based on the respiratory activity indices. This classifier is denoted as $\CLFOC$. 
For a subject with a sleep record lasting for $N$ s, the classification result of $\CLFNSA$ and $\CLFOC$ could be represented as a time series of length $2N-139$ with the range $\{\sNOR,\sOSA,\sCSA\}$, which is denoted by $s_{\text{SVM}}$.

To evaluate the accuracy of the two layer SVM classifiers composed of $\CLFNSA$ and $\CLFOC$, we reported the SVM classification accuracy of the overnight data of each subject, in which the SVM classification was trained by the same dataset.
A $3$-by-$3$ \textit{confusion percentage matrix $M$} is defined as follows:
\begin{align}
M_{i,j}=\frac{\#\{k|v_{\text{PSG}}(k)=\mathsf{i},\, v_{\text{SVM}}(k)=\mathsf{j}\}}{\#\{k|v_{\text{PSG}}(k)=\mathsf{i}\}},
\end{align}
where $i,j=\1,\2,\3$ and $\#X$ denotes the number of elements in the set $X$. Here, $\1$ indicates $\sNOR$, $\2$ indicates $\sOSA$, and $\3$ indicates $\sCSA$. Clearly, $M_{i,i}$ is the \textit{sensitivity} (SE) of the SVM classifier at the state $\mathsf{i}$. 
The specificity (SP) of the state $\mathsf{i}$ is denoted as follows: 
\begin{align}
\text{SP}(\mathsf{i}):=\frac{ \sum_{\mathsf{j}\neq \mathsf{i}}\#\{k|v_{\text{PSG}}(k)=\mathsf{j},\, v_{\text{SVM}}(k)=\mathsf{j}\}}{\sum_{\mathsf{l}}\sum_{\mathsf{j}\neq \mathsf{i}}\#\{k|v_{\text{PSG}}(k)=\mathsf{j},\, v_{\text{SVM}}(k)=\mathsf{l}\}}.
\end{align} 
The overall accuracy (AC) is defined as follows:
\begin{align}
\text{AC}:=\frac{\sum_{\mathsf{i}=\1}^{\3}\#\{k|v_{\text{PSG}}(k)=\mathsf{i},\, v_{\text{SVM}}(k)=\mathsf{i}\}}{|v_{\text{PSG}}|}.
\end{align} 
Note that these definitions are direct generalizations of the AC, SE and SP of the binary categorical response data.

\section{Database, Study Design and Results}\label{Section:Result:SVM}

\subsection{Database}

A standard PSG study was performed with at least 6 hours of sleep to confirm the presence or absence of OSA from the clinical subjects suspected of sleep apnea at the sleep center in Chang Gung Memorial Hospital (CGMH), Linkou, Taoyuan, Taiwan. The Institutional Review Board of CGMH approved the study protocol (No. 101-4968A3). Subjects with AHI$>15$ were enrolled, and the enrolled subjects provided written informed consent. THO and ABD movements were recorded by using piezo-electric bands at a sampling rate of $100$ Hz on the Alice 5 data acquisition system (Philips Respironics, Murrysville, PA). Although other standard signals such as oral-nasal airflows, electroencephalography (EEG), and oxygen saturation (SpO2) were recorded, they were not included in our analysis because we evaluated the amount of information that we could acquire from the ABD and THO signals. The exclusion criterion was the low quality of the THO and ABD movement signals. If the THO and ABD movement signals were simultaneously unrecognizable, or if the subject did not present CSA events, which were both determined by the sleep expert, the subject was excluded from the study. Forty-seven subjects were enrolled before applying the exclusion criteria and ultimately, 34 subjects were included in the analysis. The demographic details of the subjects are summarized in Table \ref{tab:SubDemo}. 

\subsection{Study design}
The sleep stages, apneas, and hypopneas were defined and scored by an experienced sleep technologist according to the AASM 2007 guideline \cite{Iber_Ancoli-Isreal_Chesson_Quan:2007}, and the scores were reconfirmed by a physician specialized in sleep medicine. On the basis of the scoring, the respiratory states during the entire night sleep were classified into five major categories: NOR, OSA, CSA, HYP, and MSA. 
As discussed previously, this study considered MSA as OSA according to clinical practice. The statistics of OSA, CSA, and HYP event numbers are summarized in Table \ref{tab:SubDemo}. Note that the number of CSAs was smaller than that of OSAs, and the number of HYPs was not negligible in this dataset.

In this study, although the features of the HYP events were distinct from that of NOR, we considered HYP as NOR and focused on classifying NOR, OSA, and CSA. In addition, for consistency with clinical practice, the ABD and THO signal segments where the sleep stage was defined as awake were excluded from the analysis. Thus, we included the ABD and THO movement signals with three respiratory states, NOR, OSA, and CSA, in this analysis. Note that we did not exclude patients with significant HYP.

\begin{table*}[t]
\centering
\caption{Demographic details of Subjects. Data are represented as ``mean $\pm$ standard deviation''.}
\label{tab:SubDemo}
\begin{tabular}{c|c|c|c|c|c|c|c}\hline
    Group      &Case  &Averaged        &Age     &BMI & CSA event & OSA event & HYP event \\	
	    	&	number	&	AHI	&	(years)	& (kg/m\textsuperscript{2})   & number & number & number \\ \hline\hline
$15<$AHI$\leq 30$ & $3$   &$21.3\pm 2.3$    &$59.5\pm 12.2$   &$24.5\pm 9.2$   & $3.4\pm3.5$  & $33\pm17.9$ & $75.1\pm 14.8$  \\\hline
$30<$AHI    & $31$   &$58.2\pm 19.8$   &$49.7\pm 12.5$   &$27.1\pm 4.6$   & $11.7\pm17.4$  & $190.7\pm122.6$ & $70.7\pm61.4$ \\\hline
\end{tabular}
\end{table*}

\subsection{Support Vector Machine Classification Result}

Based on the two features extracted from the THO signal, the overall accuracy of the two-layer SVM classifiers depending on $\CLFNSA$ and $\CLFOC$ was $75.9\%\pm 11.7\%$; the overall sensitivities of NOR, OSA, and CSA of the two-layer SVM classifiers were $73.4\%\pm 14.2\%$, $80.1\%\pm 17.7\%$, and $81.8\pm 21.9\%$, respectively; the overall specificities of NOR, OSA, and CSA of the two-layer SVM classifiers were $68.1\%\pm 11.6\%$, $73.3\%\pm 14.1\%$, and $75.9\pm 11.8\%$, respectively. 
{ When only the ABD signals were used to classify normal (NOR) or apnea (OSA+CSA) events, the overall apnea detection accuracy of the single SVM was $73.8\% \pm 4.4\%$. The overall sensitivity was $69.8\% \pm 7.8\%$ and the overall specificity was $73.6\% \pm 5.4\%$.  }

Furthermore, we reported the classification results based on four features extracted from the THO and ABD signals. After repeating the cross validation 25 times, the overall accuracy of the two-layer SVM classifiers depending on $\CLFNSA$ and $\CLFOC$ was $81.8\%\pm 9.4\%$; the overall sensitivities of NOR, OSA, and CSA of the two-layer SVM classifiers were $79.4\%\pm 9.5\%$, $88.6\%\pm 9.5\%$, and $85.4\pm 16.3\%$, respectively; the overall specificities of NOR, OSA, and CSA of the two-layer SVM classifiers were $73.9\%\pm 10.6\%$, $79.4\%\pm 9.4\%$, and $81.8\pm 9.5\%$, respectively.
Evidently, the overall accuracy, sensitivity, and specificity increased when we combined features from the THO and ABD signals. We confirmed the effect of combining features from THO and ABD on improving the overall prediction accuracy by performing the Mann-Whitney U test. Under the null hypothesis that the prediction accuracy is the same with the two features from the THO signal and the four features extracted from THO and ABD signals, we rejected the hypothesis by the Mann-Whitney U test when the p value was less than 0.02.

\section{Application to State Machine Design}\label{subsection:StateMachine}

\begin{figure}[t]
\centering
\includegraphics[width=.7\textwidth]{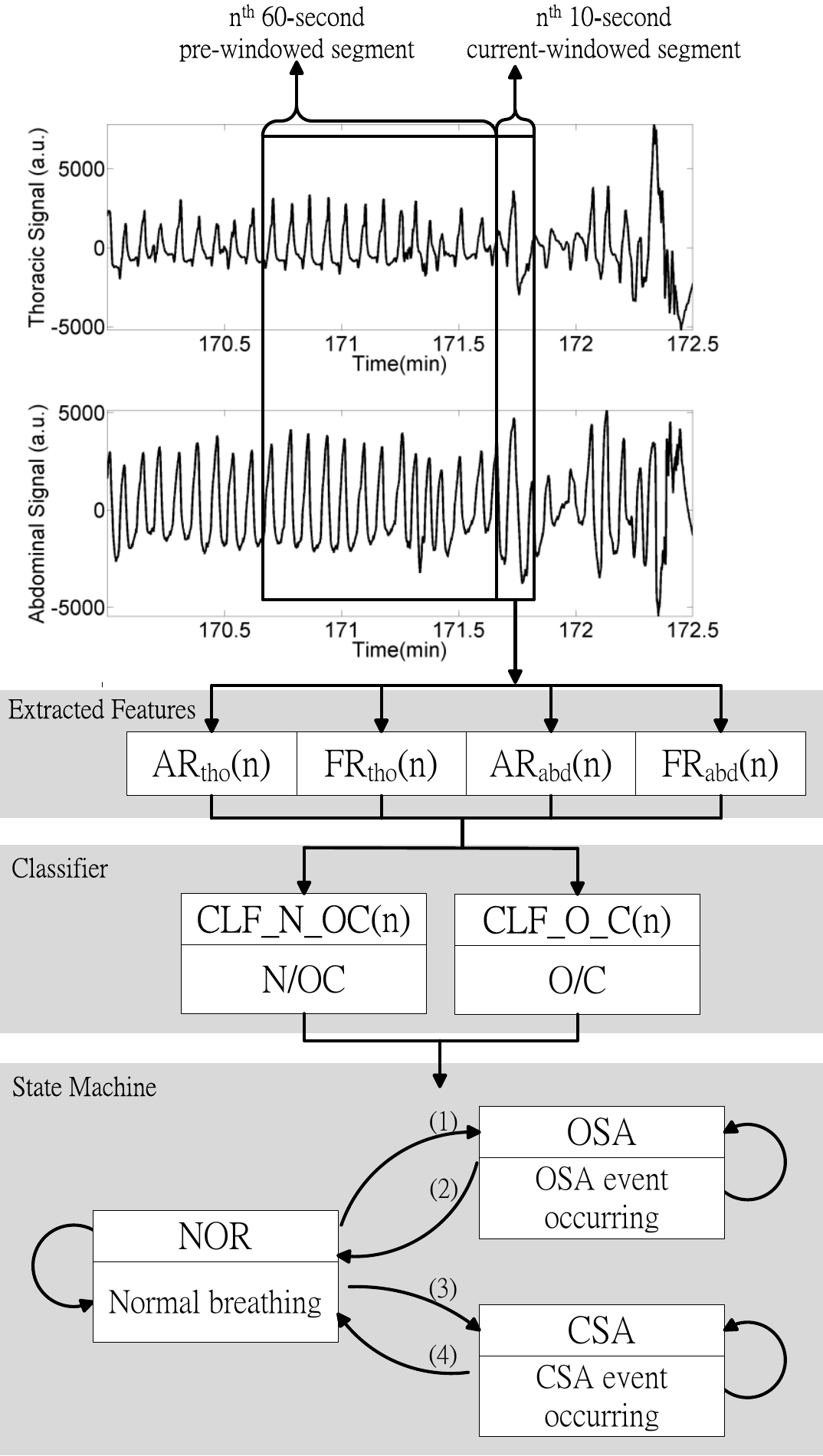} 
\caption{Flowchart of the sleep apnea event identification algorithm. Four features ($\ARtho$, $\FRtho$, $\ARabd$ and $\FRabd$)  are extracted from each segment of the thoracic and abdominal signals shown on the top panel. Here a.u. indicates arbitrary unit. The extracted features are then fed into two binary SVM classifiers, $\CLFNSA$ and $\CLFOC$, to train the SVM classifying model. The designed state machine is illustrated in the bottom panel, which contains three states, $\NOR$, $\OSA$, and $\CSA$. The transition criteria, (1)-(4), are detailed in subsection \ref{subsection:StateMachine}.} 
\label{fig:algoflow} 
\end{figure}

In this section, we demonstrate the application of the classification results from the two SVM classifiers and $\Cov$ for designing a state machine for an online prediction system. 
The flowchart of the algorithm is presented in Figure \ref{fig:algoflow}.
{Although this state machine may not improve the sleep apnea detection accuracy, it has potential applications in other sleep-related problems. For example, in a sleep lab, an online sleep apnea detection could assist in determining the optimal pressure for the continuous positive airway pressure titration; if apnea events and life-threatening arrhythmic events occur simultaneously, we could provide immediate assistance by treating the sleep apnea events, either remotely or in the sleep lab. The clinical application of the state machine will be further explored in a future study.}

We maintained the state machine as simple as possible to demonstrate the notion. The state machine was initially at the $\sNOR$ state. Figure \ref{fig:algoflow} (bottom panel) illustrates the four rules that guide the state transition. These rules depend on the binary SVM classifiers, $\CLFNSA$ and $\CLFOC$, and a chosen number of CWs, $L>0$. The number $L$ was referred to as the {\it tap number} in our study. Specifically, at the $n$-th CW, state transition occurred according to the following conditions for $\CLFNSA(n-L+1),\ldots,\CLFNSA(n)$:
\begin{enumerate}
\item[(1)] $\CLFNSA(i)=\SA$ for all $i$ and $n-L+1\leq j \leq n$ exists so that $\CLFOC(j)=\OSA$ and $\Cov(j)<0$, or at least two $l$'s exist so that $\CLFOC(l)=\OSA$;
\item[(2)] $\CLFNSA(i)=\NOR$ for all $i$; 
\item[(3)] $\CLFNSA(i)=\SA$ but (1) does not hold;
\item[(4)] the same as (2). 
\end{enumerate}
The state machine predicted sleep apnea events every $0.5$ s, on the basis of which, the location of each apnea event as well as the total number of events were outputted.

\subsection{Assessment of State Machine Event Detection Accuracy}\label{sec:AssIdx}
Although comparing the predicted event number with the ground truth event number is an essential accuracy index to evaluate the algorithm, we should also investigate whether the events are predicted in the right location; that is, a detected event should significantly overlap with a true event, and an {\it event-by-event} evaluation should also be reported. However, according to our review of relevant literature, a general consensus on the event-by-event accuracy assessment is unavailable in the field. Although developing a systematic evaluation tool to solve this problem is beyond the scope of this paper, we proposed the following two indices to evaluate the event-by-event accuracy of the proposed algorithm.

In addition to the PSG state determined by the sleep expert, the proposed state machine detection algorithm provides a predicted respiratory activity state, denoted as a time series $s_{\text{ALG}}$, with a length of $2N-139$ and range $\{\1,\2,\3\}$, representing the prediction result.
By considering $s_{\text{PSG}}$ as the ground truth, we evaluated the accuracy of $s_{\text{ALG}}$ in event-by-event detection. We define the following quantities. First, 
 
\begin{equation}
\label{eq:Def_idx_S}
S:=\frac{\sum_{i=1}^{2N-139}\delta_{s_{\text{PSG}}(i),s_{\text{ALG}}(i)}}{2N-139},
\end{equation} 
where $\delta$ is Kronecker delta function. Note that the term in the numerator indicates the number of correct state estimates by the algorithm over all segments. This quantity measures the accuracy of the states, including NOR, OSA, and CSA, estimated by the designed state machine. 

The second quantity measures the accuracy of the apnea event estimation. An {\it event period} represents a time interval over which the state is a fixed apnea type. If the respiratory activity state is OSA or CSA over an event period, we referred to the event an {\it OSA event} or {\it CSA event}, respectively. Suppose that $n_{\text{PSG},O}$ OSA events and $n_{\text{PSG},C}$ CSA events and  $n_{\text{ALG},O}$ OSA events  and  $n_{\text{ALG},C}$ CSA events are identified by sleep experts and predicted by the proposed algorithm, respectively. {Note that the AHI or AI index is not directly suitable to evaluate if the events predicted by the algorithm are accurate, since the temporal information of event periods is not taken into account in these indices. We thus need an index that could} determine whether a detected event is really an event. {To achieve this goal, we introduce the I index}:
\begin{equation}
\label{eq:Def_idx_I}
I:=\frac{P_t+A_t}{P_t+A_t+P_f+A_f},
\end{equation}
{ where $P_t$ is the sum of the number of OSA events in $s_{\text{PSG}}$ that overlaps with an OSA event in $s_{\text{ALG}}$ and the number of CSA events in $s_{\text{PSG}}$ that overlaps with a CSA event in $s_{\text{ALG}}$, { $A_t$} is the sum of the number of OSA events in $s_{\text{ALG}}$ that overlaps with an OSA event in $s_{\text{PSG}}$ and the number of CSA events in $s_{\text{ALG}}$ that overlaps with a CSA event in $s_{\text{PSG}}$, $P_f =n_{\text{PSG},O}+n_{\text{PSG},O}-P_t$ and $A_f =n_{\text{ALG},O}+n_{\text{ALG},O}-A_t$. We could view} this index {as the ``sensitivity''} of the event detection algorithm.

\subsection{Tuning the state machine}

Note that the transition rules depend on the classification results of $\CLFNSA$ and $\CLFOC$ on $L$ consecutive CWs. Evidently, in addition to the state machine structure, the tap number $L$ affects the result. To keep the state machine simple in this study, we solely tuned the tap number. We applied various tap numbers, ranging from $6$ to $20$, on the whole study population, and the tap number was optimized by evaluating the indices $I$ and $S$.  Consequently, the tap number in the proposed state machine was fixed to $L=12$.

\subsection{State Machine Detection Performance}\label{Section:Result:StateMachine}

\begin{figure*}[t]
\centering
\includegraphics[width=.85\textwidth]{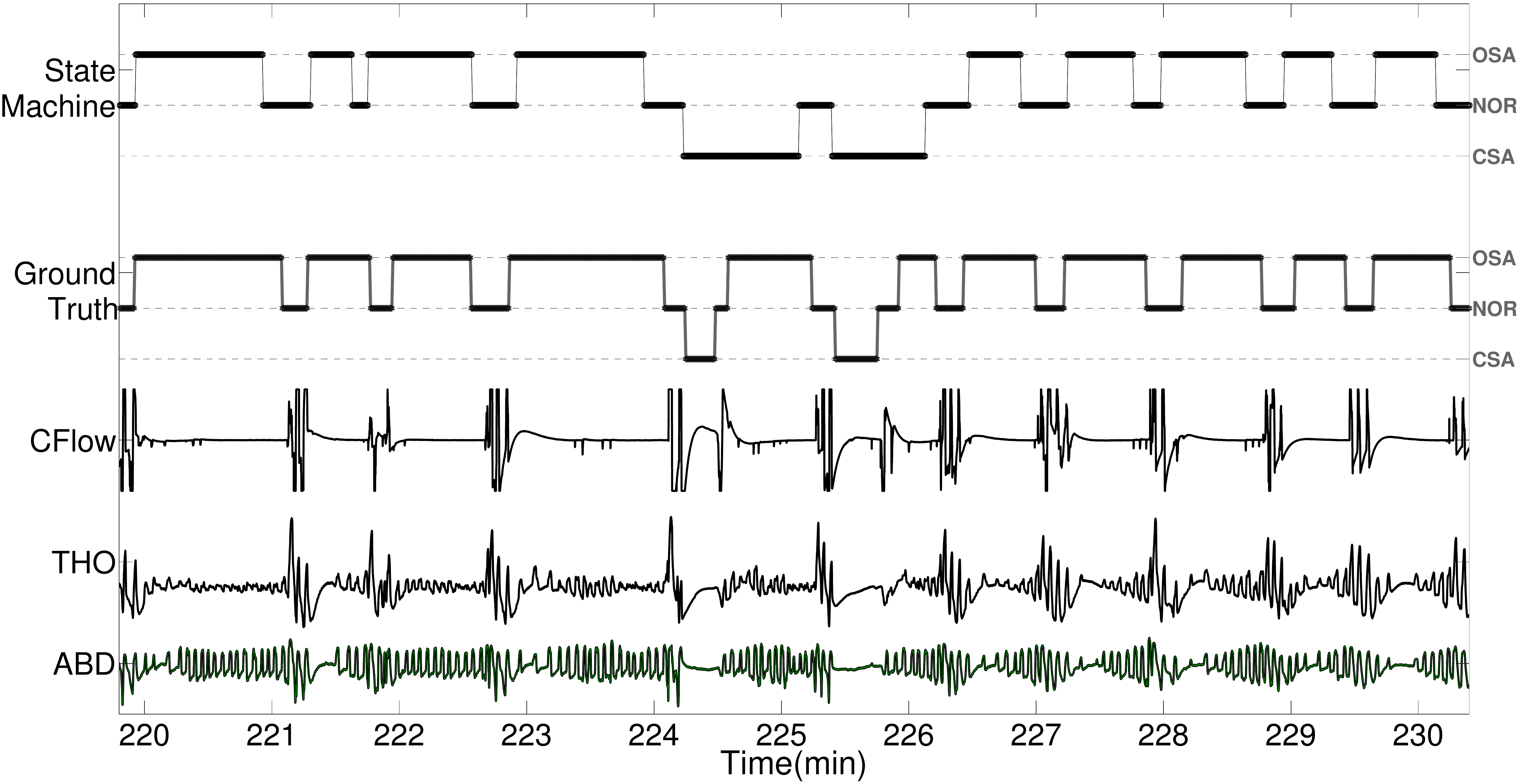} 
\caption{The event-by-event detection result obtained from the proposed on-line state machine algorithm. In the top and second panels, the sleep apnea events detected by the proposed algorithm and those determined by the sleep expert, respectively, are presented to compare the performance. Here we consider the sleep apnea events determined by the sleep expert as the ground truth. The nasal flow signal (CFlow), thoracic (THO) and abdominal (ABD) movement signals are all presented in the below panels for comparison. At various time points, although a deviation between lengths of events determined by the sleep expert and those determined using our algorithm, the events could be efficiently captured by using our algorithm.}
\label{fig:SM_rlt} 
\end{figure*}

Figure \ref{fig:SM_rlt} presents a segment of nasal flow (CFlow) and the THO and ABD signals with the detection results and the ground truth. At various time points, although a deviation between the onset and termination of events determined by the sleep expert and using our algorithm was observed, the events were efficiently captured. Note that this deviation partially originated from the existence of the group $\texttt{X}$ for the subject. Specifically, during the period in the group $\texttt{X}$, at least one sudden jump from one respiratory status to another one was observed, and this type of transition was not considered. To quantify the results, an event-by-event detection accuracy and the event number estimation accuracy indices were adopted. 

First, the S and I indices were reported. As previously stated, our algorithm did not consider HYP in the classification; however, no subject with HYP was excluded (see Table \ref{tab:SubDemo}). Thus, for an unbiased accuracy evaluation, we evaluated the accuracy under two conditions: (HYP1), by considering  HYP as NOR in the ground truth; (HYP2) by excluding the segments scored as HYP and evaluating accuracy indices on the remaining segments. The statistics of the resultant indices, S and I, in 34 subjects are summarized in Table \ref{tab:rlt_47sub}. Because HYP was considered as NOR in this study, we expected to obtain a higher accuracy under HYP2, and this expectation was fulfilled in the result. 

\begin{table}[h]
\centering
\caption{Statistics summary of the accuracy indices in 34 subjects.}
\label{tab:rlt_47sub}
\begin{tabular}{|c|c|c|c|c|}\hline
\multirow{2}{*}{} &\multicolumn{2}{c|}{Index S} &\multicolumn{2}{c|}{Index I}  \\\cline{2-5}
       &(HYP1)      &(HYP2)        &(HYP1)     &(HYP2) \\\hline
95\% quantile (\%) & 96.23    &96.97  &95.38      & 95.61        \\\hline
Median (\%)  & 81.93    & 84.08  & 66.84     & 82.61       \\\hline
5\% quantile  (\%) & 69.83    & 70.59 &25.98      & 40.17        \\\hline
Mean (\%)  & 82.4    & 84.01  &67.36      & 77.21       \\\hline
Standard Deviation (\%)  &  8.89   & 9.06 & 22.06     &  19.01  \\\hline
\end{tabular}
\end{table}

In addition to the event-by-event accuracy indices, the detected event number is crucial for screening the severity of the condition in a subject. Thus, the accurate AHI estimation is usually expected in the literature. However, because we did not consider HYP in this study, we reported the estimated number of the apnea events, which could be considered as an index parallel to AI. The accuracy of the event number detection is defined by $100\%\times\left(1-\frac{|\text{true number}-\text{estimated number}|}{\text{true number}}\right)$. The accuracy of the event number detection was $73.46\pm 18.26\%$ under the HYP1  condition and $84.42\pm 11.24\%$ under the HYP2 condition.

\section{Discussion}\label{sec:Disc}
 
This study explored the possibility of not only identifying sleep apnea but also distinguishing OSA from CSA by carefully designing features and classifiers for a single THO signal. Moreover, by including the features extracted from the ABD signal, we could obtain a more favorable result. In addition to providing a mathematical model to quantify the intrinsic features within the respiratory signals and applying SST to reduce the noise influence in an on-line fashion, we proposed suitable features hidden in the THO and ABD signals to capture OSA and CSA. SVM was applied to establish a classifier for the apnea events. The designed features and the SVM classifier were applied to design a state machine for a potential online apnea detection algorithm. Furthermore, two event-by-event accuracy indices are proposed to further evaluate the performance of the proposed algorithm. 
The useful information hidden inside the THO or ABD signals leads to the possibility of designing an easy-to-install, non-invasive, and non-intrusive level IV sleep apnea detection equipment (the AASM criteria) by combining THO and one more potential sensor. 

In this study, to mimic complicated real world problems, our exclusion criteria for data collection were quite stringent. Certainly, because we only excluded subjects with CSA and OSA signal qualities too low to be identified even by sleep experts, several subjects in our database have only one recognizable THO or ABD movement signal. Because we did not remove these cases, clearly, the results from the SVM classifier were downgraded by these cases. In addition, it is well known that the SVM performance is affected when the sizes of the two groups under classification are markedly distinct. Although this effect was corrected in the SVM algorithm, the negative impact of the uneven distributions of CSA and OSA cannot be ignored.  
Under such conditions, however, we still obtained a satisfactory result from the SVM classifier after cross validation, which confirms the robustness and suitability of the proposed features extracted from the THO and ABD signals.

Compared with the SVM classification result, initially, the state machine results do not appear to be more favorable than most of the reported results. We should, however, note that the evaluation standards for the state machine application are different - the event-by-event accuracy indices are our main indices. In addition, the inter-observer event disagreement problem exists -- the mean agreement rate among various scorers, even in the normal subjects, is 76\% with a range of 65-85\%. A similar disagreement in apnea scoring was also reported in \cite{Bridevaux_Fitting_Fellrath_Aubert:2007} among observers -- the intraclass correlation coefficients were $0.73$ for agreement on AHI and $0.71$ for hypopnea index. Therefore, obtaining an event-by-event accuracy higher than $80\%$ may not be meaningful. 
Thus, the state machine result obtained in our study is satisfactory and the proposed algorithm has potential for screening in various sleep apnea patients (OSA v.s. CSA) with various pathophysiological mechanisms (upper airway obstruction vs. ventilation control instability).

\begin{figure}[t]
\centering
\includegraphics[width=.85\textwidth]{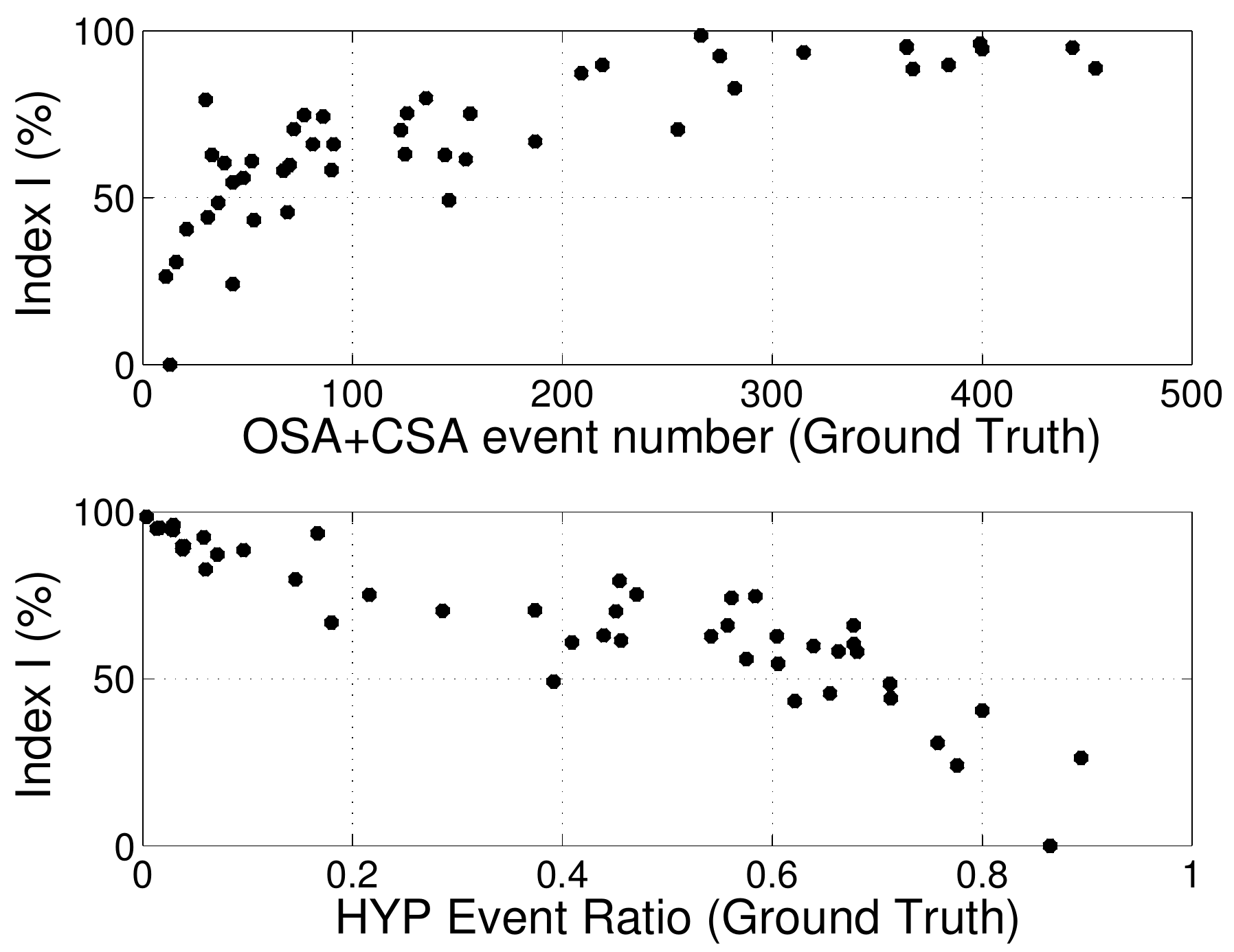} 
\caption{The relation between the number of the hypopnea (HYP) events that occur in a subject and the accuracy index I obtained by applying the proposed algorithm to the thoracic and abdominal signals of the subject. The index I is evaluated under condition HYP1; that is, HYP events are considered as NOR in the ground truth. HYP event ratio is derived by dividing the total number of HYP events by the total number of OSA, CSA, and HYP events.} 
\label{fig:Evid_I_v1} 
\end{figure}

Our study has some limitations. %
In particular, the problem of the HYP events should be discussed. In this study, we did not distinguish the hypopnea and non-hypopnea events in the NOR group because the proposed features could not efficiently distinguish HYP (lower than 50\% by the SVM classifier). Furthermore, the HYP events could downgrade the final analysis result.
To further visualize the influence of the HYP events on the analysis, the number of HYP events versus the index I is shown in Figure \ref{fig:Evid_I_v1}. With the significance level set as $0.05$, the linear relationship between the number of HYP events and index I was significant. 
To resolve this limitation, we should design new features by considering the THO (or the ABD) piezo-electric sensor, the distinguishable phenomenological and physiological characteristics of HYP, and their relationship. Another solution is combining the THO signal with one more channel, such as the oximeter, which {provides the essential information for HYP diagnosis \cite{Iber_Ancoli-Isreal_Chesson_Quan:2007,Berry_Budhiraja_Gottlieb:2012}.} The result will be reported in the future.
Second, we did not design features to capture the transition periods, {that is, the events} in group $\texttt{X}$ in the training process, which limited the accuracy of the testing result. {Precisely, because the information contained in the transition periods {is} a combination of two breathing patterns, combining a regression technique might help to improve the {result.}} 
Third, the low quality of the recorded THO {and/or} ABD signals must be resolved by adopting a more accurate sensor or signal recording method. {{ An alternative solution is incorporating} the signal quality index (SQI) {into} the design. Although several SQIs are available for other signals \cite{Karlen2012,Orphanidou2015}, a suitable SQI for the THO and/or ABD signals is {less explored}.}
Fourth, although MSA could be practically considered as OSA, its dedicated physiological implication should not be ignored. Because MSA is a complex combination of CSA and OSA, a more sophisticated algorithm with more dynamical features is required to distinguish MSA. {Particularly, we could {design} features by considering the physiologically special breathing pattern structure during MSA.}

\subsection{Future works in progress}

First, under suitable conditions, the current simple state machine could be replaced with a more sophisticated classification model such as the neural network \cite{Khandoker_Gubbi_Palaniswami:2009}.
Second, the inter-individual discrepancy must be resolved for the screening purpose. 
Third, on the basis of the preliminary study reported in \cite{Wu_Talmon_Lo:2015}, an automatic awake-sleep status detection algorithm will be incorporated into the algorithm for the screening purpose. 
{Fourth, starting from 2012, the respiratory inductance plethysmography (RIP) sensor became a recommended sensor for the sleep examination by AASM \cite{Berry_Budhiraja_Gottlieb:2012}. 
Since the RIP sensor depends on an inductive coil whose electromagnetic properties are related to the area enclosed, it may suffer less from the trapping artifact commonly seen in the piezoelectric sensor  \cite{Butkov_Lee-Chiong:2007}, and could provide another dimension of respiratory information \cite{Martinot-Lagarde1988}. The proposed model and analysis tools might extract more information from the RIP signal and help extract information more accurately \cite{DeGroote2000}. Because the above problems are beyond the scope of this paper, a systematic study will be reported in a future paper. 
Fifth,} based on the advances in the chip design technique, we could include various non-invasive sensors, such as the three-axis accelerometer and wireless radar sensor, in a portable device. These non-invasive and non-intrusive sensors not only provide more information for identifying sleep apnea and distinguishing OSA, CSA, and HYP, but also greatly reduce the interference in the sleep, thereby increasing the reliability of overnight sleep testing. Based on the results demonstrated in this paper, we could incorporate various signals to get a more favorable prediction result. 

\section{Conclusion}\label{sec:Conc}

This study extensively explored the capability of solely using the THO signal and combining THO and ABD signals to detect the OSA and CSA events through wearable piezo-electric bands, which has been rarely investigated in previous studies. The evaluation of the proposed algorithm yielded satisfying results by using the cross-subject validation procedure. Moreover, the results verified the competency of the selected features in distinguishing between the OSA and CSA events. Despite the limitations discussed in Section~\ref{sec:Disc}, this study laid the foundation for using THO and/or ABD  signals in sleep apnea detection as well as presented optimistic potential of applying the proposed algorithm to the clinical examinations for both screening and homecare purposes.

\section*{Acknowledgement}
Hau-tieng Wu acknowledges the support of Sloan Research Fellowships, FR-2015-65363. This work was supported by the Ministry of Science and Technology
(MOST), Taiwan, under grant number MOST 103-2220-E-007-009.

\bibliographystyle{amsplain}
\bibliography{sleep}

\end{document}